\documentclass[sigconf]{acmart}

\usepackage{balance}

\AtBeginDocument{%
  \providecommand\BibTeX{{%
    \normalfont B\kern-0.5em{\scshape i\kern-0.25em b}\kern-0.8em\TeX}}}

\copyrightyear{2023} 
\acmYear{2023} 
\setcopyright{acmlicensed}\acmConference[WebSci '23]{15th ACM Web Science Conference 2023}{April 30-May 1, 2023}{Evanston, TX, USA}
\acmBooktitle{15th ACM Web Science Conference 2023 (WebSci '23), April 30-May 1, 2023, Evanston, TX, USA}
\acmPrice{15.00}
\acmDOI{10.1145/3578503.3583596}
\acmISBN{979-8-4007-0089-7/23/04}

\usepackage{multirow}
\begin{document}

\title[Geolocated Social Media Posts are Happier]{Geolocated Social Media Posts are Happier:
Understanding the Characteristics of Check-in Posts on Twitter}
\author{Julie Jiang}
\email{juliej@isi.edu}
\affiliation{
\department{Viterbi School of Engineering}
\department{Information Sciences Institute}
\institution{University of Southern California}
 \country{{USA}}}

 \author{Jesse Thomason}
\email{jessetho@usc.edu}
\affiliation{
\department{Information Sciences Institute}
\institution{University of Southern California}
 \country{{USA}}}
 
\author{Francesco Barbieri}
\email{fbarbieri@snap.com}
\affiliation{\institution{Snap Inc.}
\country{USA}}

\author{Emilio Ferrara}
\email{emiliofe@usc.edu}
\affiliation{
\department{Annenberg School of Communication}
\department{Viterbi School of Engineering}
\department{Information Sciences Institute}
\institution{University of Southern California}
 \country{{USA}}}



\begin{abstract}
The increasing prevalence of location sharing features on social media has enabled researchers to ground computational social science research using geolocated data, affording opportunities to study human mobility, the impact of real-world events, and more. This paper analyzes what crucially separates tweets with geotags from tweets without. Our findings show that geotagged tweets are not representative of Twitter data at large, limiting the generalizability of research that uses only geolocated data. We collected 1.3M geotagged tweets on Twitter (most of which came from Instagram), and compared them with a random dataset of tweets on three aspects: affect, content, and audience engagement. We show that geotagged tweets on Twitter exhibit significantly more positivity, often citing joyous and special events such as weddings, graduations, and vacations. They also convey more collectivism by using more first-person plural pronouns and contain more additional features such as hashtags or objects in images. However, geotagged tweets generate less engagement. These findings suggest there exist significant differences in the messages conveyed in geotagged posts. Our research carries important implications for future research utilizing geolocation social media data. 
\end{abstract}





\maketitle

\section{Introduction}

Location sharing features on social media have become prevalent following advancements in smartphones location services, powering researchers to study topics such as patterns of 
human mobility or activity ~\cite{noulas2011empirical,botta2015quantifying,jurdak2015understanding,hasan2016understanding,rashidi2017exploring,xu2020twitter,yang2021online,phillips2021social}, 
effects of weather and climate on sentiment~\cite{coviello2014detecting,dahal2019topic,wang202043,jiang2022sunshine}, natural disasters~\cite{de2009omg,crooks2013earthquake}, disease outbreaks~\cite{padmanabhan2014flumapper,wang2022global}, and political campaigns~\cite{hobbs2019effects,yaqub2018analysis,wang2022global}.

Despite the abundance of literature utilizing geolocation to study large-scale human behavior, users who geotag, or check-in,\footnote{We use the terms geotag and check-in interchangeably.} are not representative of the social media user population~\cite{sloan2015tweets,huang2019large,karami2021analysis}.
This inconsistency could undermine the usefulness and validity of research leveraging geolocated data. Some literature studied user motivations for using geolocation services~\cite{lindqvist2011im,patil2012check,patil2012reasons,guha2013can,koohikamali2015location,tasse2017state} and others tapped into demographic, cultural, and linguistic choices between those who geotag and those who do not~\cite{sloan2015tweets,rizwan2018using,huang2019large,karami2021analysis}. However, to the best of our knowledge, less is understood about the differences between check-in posts and non-check-in posts.

To bridge the gap in understanding geolocation correlates on social media, we examine aspects that distinguish check-in tweets from non-check-in tweets. Our research is carried out in three broad directions: (1) the affective state of the tweet, characterized by the expressed sentiment (valence) and emotion, (2) the textual and non-textual content characteristics of the tweet, and (3) the engagement generated by the tweet on Twitter as a proxy for popularity and influence. Using 1.3 million check-in tweets, most of which came originally from Instagram, and a random dataset of tweets of the same size, we answer the following research questions:
\begin{itemize}
    \item \textbf{RQ1: Do check-in tweets exhibit higher positivity bias in sentiment and emotions?} We find that check-in tweets express more positive sentiments and joyful emotions.
    \item \textbf{RQ2: Are there differences in content between check-in and non-check-in tweets?} We observe several distinguishing characteristics of check-in and non-check-in tweets in terms of both textual and non-textual content. Check-in tweets are often about special and positive occasions (e.g., weddings) or tourist destinations. They also convey more collectivism with first-person plural pronouns and contain more features such as hashtags.
    \item \textbf{RQ3: Are there differences in audience engagement between check-in and non-check-in tweets?} We show that check-in tweets receive less engagement in terms of likes, retweets, etc.
\end{itemize}


In spite of its limited capacity to be representative of general social media data, geolocated social media data remains valuable resources for researchers and decision-makers. This research sheds light on the crucial differences between geolocated and non-geolocated social media data from vantage points of affect, content, and engagement. Our findings can inform future studies using geolocation data their unique characteristics and potential to generalize.

\section{Background and Related Work}
\subsection{Geotagging on Social Media}

The availability of location information in social media has empowered research studying the relationship between user location and behavior social network platforms such as Twitter (e.g., ~\cite{jurdak2015understanding,phillips2021social}) but also Facebook~\cite{coviello2014detecting}, Snapchat~\cite{yang2021online,jiang2022sunshine}, Weibo~\cite{wang202043}, and Four- square~\cite{cho2011friendship,noulas2011empirical,hasan2016understanding}. Existing work falls broadly into three categories: studying real-world phenomena using social media geolocation, understanding motivations behind geotagging usage, and characterizing users who geotag.









\subsubsection{Computational Social Science via Geotags.} 
There is a vast body of literature utilizing geotags to empirically study real-world phenomena. Many of these studies use geolocation to track human mobility~\cite{noulas2011empirical,conover2013geospatial,jurdak2015understanding,botta2015quantifying,rashidi2017exploring,phillips2021social} and understand the dynamics of social influence~\cite{cho2011friendship,monsted2017evidence,phillips2021social}, communications~\cite{yang2021online}, or trends \cite{ferrara2013traveling}. \citet{gu2016twitter} utilized geotagged tweets to detect real-time traffic incidents~\cite{gu2016twitter}. By combining geolocation with content understanding, researchers have analyzed the public's response to elections~\cite{yaqub2018analysis}, political campaigns~\cite{hobbs2019effects}, or the COVID-19 pandemic~\cite{wang2022global}. Moreover, geolocation can help examine the effects of naturally occurring events or disasters, such as the effects of weather on mood~\cite{coviello2014detecting,jiang2022sunshine}, attitudes toward climate change~\cite{dahal2019topic,wang202043}, the detection of earthquake impact areas~\cite{crooks2013earthquake}, the spatial analysis of forest fires~\cite{de2009omg}, or the spread of disease outbreaks~\cite{padmanabhan2014flumapper}. \citet{gupta2013faking} leveraged location data to discern fake images of hurricane Sandy from real ones. Geolocation also helped \cite{larson2019social} analyze social network structures in protest participation. A separate line of research predicts user locations from geotags~\cite{ajao2015survey}. 

The extensive literature that concerns geolocation data in computational social science underscores the need to understand how geotagged social media posts are distinct. Most research limit their findings to users who geotag, missing the opportunity to reasonably generalize their work to larger body of users who do not geotag. By understanding \textit{how} geotagged posts separate from non-geotagged posts, this work provides a path towards generalizing research concerning only geotagged posts to the broader population.

\subsubsection{Reasons for Geotagging.} Various explanations have been posited to explain why users geotag. The majority of Twitter users geotag consciously and intentionally to communicate where they are as a socializing mechanism~\cite{lindqvist2011im,patil2012check,patil2012reasons,guha2013can,tasse2017state}. Importantly, several studies cite the importance of location-sharing as a means of impression management by portraying oneself in an interesting light ~\cite{lindqvist2011im,patil2012check,patil2012reasons,guha2013can}. \citet{guha2013can} additionally suggest that tendencies to share location depend on the visibility of the shared location to the social network.
Some  users are motivated to geotag due to incentives or rewards, such as receiving virtual badges or offline discounts~\cite{lindqvist2011im,patil2012reasons,patil2012check,koohikamali2015location}. 

Chief among the reasons for \textit{not} sharing location is privacy. Users tend to eschew divulging private information in regard to their location~\cite{tsai2010location,chang2014college,tasse2017state,hsieh2020traces}. The perceived social benefits and usefulness, along with subjective norms and attitudes towards geotagging, must outweigh concerns of privacy for users to enable geotagging~\cite{chang2014college,hsieh2020traces}. Indeed, most people only geotag rare places such as vacation spots they visit~\cite{tasse2017state} rather than locations close to home. 

The majority of these studies analyzed platforms such as Four-
square that were designed specifically for location sharing~\cite{guha2013can,lindqvist2011im,cho2011friendship}. Twitter, in contrast, is not primarily a location sharing platform, which makes it more appropriate to study in this paper the characteristics of the rare tweets that are geotagged, in comparison with general, non-geotagged tweets.





\subsubsection{Characteristics of Users Who Geotag.} Existing work argues that users who geotag are not representative of all Twitter users~\cite{sloan2015tweets,rizwan2018using,huang2019large,karami2021analysis}. For instance, users who geotag are more likely to be male and slightly older~\cite{sloan2013knowing}. \citet{rizwan2018using} also found similarly on Sina Weibo that there exist gender differences between users who geotag and uses who do not. Those who geotag also have distinct language uses ~\cite{sloan2015tweets,huang2019large}. \citet{huang2019large} further suggest that geotagging users demonstrate homophily in that they tend to be friends with those who also geotag. There are also linguistic and sharing differences between users who do and do not geotag, with geotagging users more likely to share original tweets rather than retweets, use more personal pronouns in their text, and use more hashtags~\cite{karami2021analysis}. Personality traits such as extroversion-introversion are also associated with a user's proclivity to geotag~\cite{wang2013share}. However, despite the research presented above for geotagged tweets, gaps exist in understanding the differences in affect, content, and influence between geotagged and non-geotagged tweets.

\subsection{Background on Twitter}

Twitter is a microblogging platform where people can post short, original updates (``tweets'') or share (``retweets''), like, and comment on other people's tweets in real-time.

\subsubsection{Affect in Tweets.}
Research on affect in tweets is varied, ranging from slightly more positive \cite{lerman2018language}, to mostly neutral \cite{ferrara2015measuring,ferrara2015quantifying}, to mostly negative even when it pertains to positive events \cite{naveed2011bad,thelwall2011sentiment,ferrara2015measuring}. \citet{murthy2015we} found that tweets from mobile sources are 25\% more negative than those from web-sources. However, no works indicate that the Twitter affect is largely positive. \citet{waterloo2018norms} suggests that Twitter users feel more appropriate to share negative emotions on and less appropriate to share positive emotions on Twitter than on other platforms such as Instagram. The platform norms could be attributed to the design of Twitter, characterized by nonreciprocal following, mainly public audience, and succinct commentaries in real-time.


\subsubsection{Geotagging on Twitter.}
Twitter users can add location information to their original tweets~\cite{twitterlocfaq}, though less than 1\% of all tweets are geotagged~\cite{ajao2015survey}. This tweet-level location tagging is not to be confused with the user-provided profile locations, which are much more readily available but less formalized~\cite{hecht2011tweets,sloan2013knowing,jiang2020political}, ranging from broad approximations (e.g., ``California'', ``Seattle'') to fictitious locations (e.g., ``somewhere over the rainbow''). Additionally, there is a fundamental difference between geotagging and using profile locations, with the former likely disclosing where the tweet is composed and the latter likely disclosing where the user lives~\cite{sloan2015tweets}. In this work, we focus on tweet-level geotagging with verifiable geolocations~\cite{twitterlocfaq}.

\subsubsection{Our Work:} Prior work studying geotagging on social media faces several limitations. Most do not compare check-in posts with non-check-in posts or study only differences between \textit{users} who geotag versus those who do not~\cite{sloan2015tweets}. Most similar to our work is \citet{karami2021analysis}; however, they reveal only that geotagged content is different from non-geotagged posts but not \textit{how} they differentiate. In this work, we take a closer look at the affect expressed, message conveyed, and engagement generated by check-in posts compared with those of non-check-in posts.

\section{Data}\label{sec:data}
We collect two datasets comparable in size and timeframe, one limited to only check-in tweets and the other a random selection of tweets. The \texttt{Check-In} dataset consists of 1.3 million geotagged original (i.e., not retweets, replies, etc.) English tweets from October 2015 to February 2017 located in 30 major US and Canadian cities. To match the \texttt{Check-In} data, we collect the \texttt{Random} dataset consisting also of 1.3 million random English original tweets pulled from a similar timeframe spanning the year 2016. The \texttt{Random} dataset therefore represent the typical selection of Tweets one might post on Twitter, regardless of influence, location, or topic. The \texttt{Random} tweets are collected with a substring matching search query that matches any of the 26 letters of the English lowercase alphabet. To ensure randomness in the collection time, which does not come natively through the Twitter API, we build an array of every unique timepoint accurate to the seconds and randomly sample a timepoint during our predefined time period for every batch of tweets we pull. The batch size is fixed to 600, which is the maximum number of tweets we can pull at once given our API quota. 

For every tweet, we collect the tweet text, user description, and the source of the tweet. When available, we also collect the geolocation and media attachments. There are three types of media attachments: images, videos, and GIFs. The video and GIFs are provided as static image renderings.
\begin{table}
    \centering
    \begin{tabular}{llll}
        \toprule
         &&  \texttt{Check-In}  & \texttt{Random} \\
         \midrule
         \multicolumn{2}{l}{\textbf{No. Tweets}} & 1,324,900 & 1,354,061 \\
         \multicolumn{2}{l}{\textbf{No. users}} &418,374 &1,053,726 \\
        \midrule
        \multicolumn{2}{l}{\textbf{Source}} \\
        & Twitter & 4\% & 54\% \\
        & Instagram & 91\% & 6\% \\
         \midrule
         \multicolumn{2}{l}{\textbf{Mobile Source}} & $99\%$ & $50\%$\\
         \midrule
        \multicolumn{2}{l}{\textbf{Scraped Media}} & 0.5\% (6,173) & 14\% (193,849) \\
        \midrule
        \multicolumn{2}{l}{\textbf{Geotagged}} & 100\% & 6\% (80,935)\\
         \bottomrule
    \end{tabular}
    \caption{Summary of key data statistics.}
    \label{tab:data_stats}
\end{table}

There are several key differences between the two datasets (Table \ref{tab:data_stats}). First, there are significantly fewer users in the \texttt{Check-In} data than in the \texttt{Random} data, owing to the fact that random tweets are sampled from a much larger pool of users. We also note that although the \texttt{Check-In} dataset was pulled from Twitter, the vast majority (91\%) of the tweets actually originated from third-party platforms, namely Instagram. This observation is in line with previous literature on the sources of geotagged tweets~\cite{hu2020understanding}.
In addition, more than 99\% of sources from \texttt{Check-In} are mobile sources, compared to around 50\% from \texttt{Random}. A more detailed breakdown of top sources can be found in the Appendix. These numbers suggest that there may be fundamental platform use differences between check-in users and non-check-in users. We discuss this issue further in \S\ref{sec:limitations}. Finally, despite the overwhelming prevalence of  \texttt{Check-In}  posts originating from Instagram, a platform primarily centered on photo and video sharing, we were only able to pull 0.5\% of the attached media, as opposed to 14\% from \texttt{Random}. This is not because \texttt{Check-In} posts had fewer media but rather the media does not transfer to the Twitter API. As pulling media from Instagram faces privacy and technical barriers, we are unable to obtain them.

The dataset is recollected in late 2021 for two reasons. First, this enables us to collect the engagement metric counts (e.g., retweet counts) well after the time of posting, approximating the eventual engagement generated by the posts. Additionally, in accordance with Twitter's API policy, this ensures that we exclude any tweets that are suspended, removed, modified, or otherwise withheld, either due to tweet deletion or user account removal \cite{twitterpolicy}.

It is worth mentioning that Twitter's geolocation service has undergone considerable redesigns over the years. Most notable is a UI change in 2019 which removed precise geotagging in place of more simplified and less precise geotags~\cite{niemanlab}. We utilize data from 2016, before this UI change. We do not believe this significantly alters the validity of the conclusions we draw as they apply today since this API change does not affect tweets that were posted to Twitter via third-party applications such as Instagram. Since Instagram-sourced tweets make up the vast majority of our \textit{Check-In} data, it should not change the nature of check-in posts on Twitter ~\cite{hu2020understanding}.

\section{Methods}
\subsection{Sentiment and Emotion Detection}

To answer \textbf{RQ1}, we use sentiment and emotion detection models \texttt{twitter-roberta-base-sentiment} and
\texttt{
\hyphenchar\font=`\- 
     \hyphenpenalty=10000 
     \exhyphenpenalty=-50 
     twitter-roberta-base-emotion}, respectively, both trained with RoBERTa-base on a Twitter corpus \cite{barbieri2020tweeteval}. They achieved competitive results on benchmark Twitter evaluation tasks \cite{barbieri2020tweeteval}. The sentiment classifier predicts sentiments of \textit{negative}, \textit{neutral}, and \textit{positive}. The emotion classifier predicts one of four emotions: \textit{joy}, \textit{optimism}, \textit{sadness}, and \textit{anger}. 

\subsection{Tweet Content Characterization}

To answer \textbf{RQ2}, we employ a mix of text and image processing techniques to characterize the differences between \texttt{Check-In} and \texttt{Random} tweets.

\subsubsection{Top Unigrams and Emojis.} As a first step towards content understanding, we determine words (unigrams) and emojis that appear relatively more frequently in one dataset. For this, we use the weighted log-odds method with uniform Dirichlet priors developed by \citet{monroe2008fightin} for lexical feature selection.

\subsubsection{Event Detection.}
An additional method to determine differences in content is through event detection. To this end, we apply the state-of-the-art event detection pipeline EventPlus~\cite{ma2021eventplus} to the tweets. EventPlus can detect the word that acts as the event trigger---usually a verb---as well as the \textit{type} of the event. The event types include \textit{Movement}, \textit{Life}, \textit{Contact}, \textit{Conflict}, \textit{Personnel}, \textit{Transaction}, \textit{Business}, and \textit{Justice}. Following \citet{sun2021men}, we define events by their corresponding trigger words and compute the odds ratio~\cite{szumilas2010explaining} of an event $e_i$ occurring in \texttt{Check-In} as opposed to in \texttt{Random}. Let $\mathcal{E}^c$ and $\mathcal{E}^r$ be dictionaries mapping events to their frequency of occurrence in \texttt{Check-In} and \texttt{Random} respectively. The odds ratio for event $e_i$ is therefore determined by:
\begin{equation}
    OR(e_i) = \frac{\mathcal{E}^c(e_i)}{\sum_{j\neq i} \mathcal{E}^c(e_j)} /  \frac{\mathcal{E}^r(e_i)}{\sum_{j\neq i} \mathcal{E}^r(e_j)}.
\end{equation}
An odds ratio of 1 indicates that the event is equally likely to occur in either dataset. The larger the odds ratio, the more likely the event occurs in \texttt{Check-In}.

For each event category, we rank trigger words by their odds ratios and take the top 5 triggers, which are most associated with \texttt{Check-In} tweets, as well as the bottom 5 triggers, which are most associated with \texttt{Random} tweets. Infrequent triggers that appeared fewer than 10 times in total are removed. 

\subsubsection{Tweet Features.} We utilize several ways to compare \texttt{Check-In} and \texttt{Random} features, including the average number of characters (length of the post), tokens, mentions (using `@'), hashtags, and URLs used in both datasets. To understand tweets on the dimension of individualism-collectivism \cite{twenge2013changes}, we also count the number of times first-person singular pronouns (`I',  `me', `my', `myself', `mine') and  first-person plural pronouns (`we', `us', `our', `ours', `ourself', `ourselves') appear in both datasets.

\subsubsection{Image Processing.}
We extract image features using the state-of-the-art object detection algorithm Faster R-CNN~\cite{ren2015faster}, which is trained to detect 20 types of everyday objects such as \textit{persons}, \textit{chairs}, \textit{monitors}, etc. We use the Faster R-CNN pretrained on \texttt{
\hyphenchar\font=`\- 
\hyphenpenalty=10000 
\exhyphenpenalty=-50 
resnet50\_v1b\_voc}~\cite{gluoncvnlp2020}.

\subsection{Engagement Metrics}
To answer \textbf{RQ3}, we compare various metrics of popularity and influence through retweet counts, mention counts, reply counts, and like counts. Since the engagement metrics returned by the Twitter API reflect the number of retweets, mentions, replies, and likes at the time of collection, and since we collected the data well after the original time of posting, we are confident that our collected metrics approximate the true eventual engagement the tweets generate. 

\subsection{Data Type Prediction}
To further rigorously examine the differences between \texttt{Check-In} and \texttt{Random} tweets, we use the aforementioned features to predict whether a tweet comes from the \texttt{Check-In} or \texttt{Random} dataset. Specifically, we group the features into four categories:
\begin{enumerate}
    \item \textbf{Sentiment and emotion:} the predicted sentiment and emotions scores from the tweet text using \texttt{twitter-roberta} finetuned for sentiment and emotions, respectively~\cite{barbieri2020tweeteval}.
    \item \textbf{Content (textual):} the pooled output embeddings of the tweet text using \texttt{BERTweet-BASE}~\cite{bertweet}.
    \item \textbf{Content (non-textual)}: the detected objects in images using Faster-RCNN~\cite{ren2015faster} and also  the number of hashtags, accounts mentioned, URLs, characters, and tokens used.
    \item \textbf{Engagement:} the number of retweets, replies, quotes, and likes the tweet generated.
   
\end{enumerate}
\begin{figure}
    \centering
    \includegraphics[width=\linewidth]{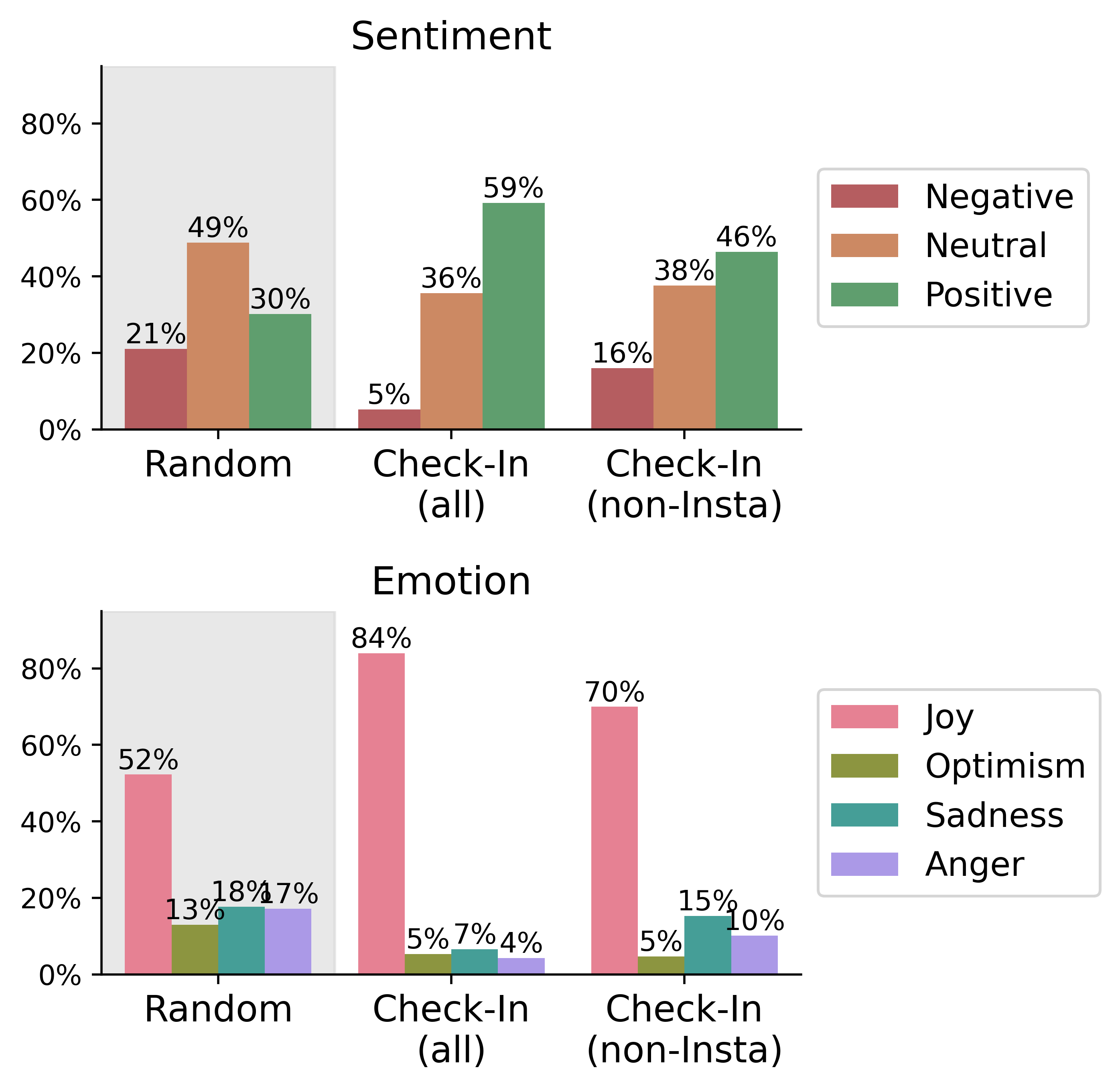}
    \caption{The distributions of detected sentiment (top) and emotions (bottom) show that \texttt{Check-In} tweets exhibit significantly more positivity than \texttt{Random} tweets.}
    \label{fig:sent_emo}
\end{figure}
The inputs are concatenated before being fed into a deep neural network for binary classification. For each combination of features, we conduct 10 runs of randomized hyperparameter tuning on five-fold cross validation. We then repeat the best hyperparameter combination with 10 different random seeds to ensure the consistency of the results and to conduct significance testings. The hyperparameter grid we used is in the Appendix.


\section{Results}

\subsection{RQ1: Positivity Bias}

We show the distributions of the detected sentiment and emotions labels in Figure \ref{fig:sent_emo}. All categorical distributions are significantly different between the \texttt{Check-In} and the \texttt{Random} datasets using a Chi-squared test ($p<0.001$). Compared to \texttt{Random} tweets, there exists significant positivity bias in \texttt{Check-In} tweets both in terms of sentiment and emotions. The proportion of tweets labeled \textit{positive} in the \texttt{Check-In} dataset is almost twice as much as that of \texttt{Random} tweets.  Correspondingly, there are significantly fewer tweets labeled \textit{negative}. In terms of emotions, \texttt{Check-in} tweets are also predominantly (84\%) labeled \textit{joy}, as opposed to \textit{sadness} or \textit{anger}. The positivity bias in sentiment and emotions remains significant even when Instagram-sourced  \texttt{Check-In} posts are excluded.
\begin{table*}
    \centering
    \begin{tabular}{cclcl}
    \toprule
    \multicolumn{1}{c}{\textbf{Event}}  & 
    \multicolumn{2}{c}{\textbf{\texttt{Check-In}}} &  
    \multicolumn{2}{c}{\textbf{\texttt{Random}}}\\
    \cmidrule(lr){1-1} 
    \cmidrule(lr){2-3} \cmidrule(lr){4-5}
    \multirow{2}{*}{\textit{Life}}  & 
    \multirow{2}{*}{27\%} &  weddingseason, weddingday, killem,  &
    \multirow{2}{*}{27\%} & assaulting, assault, raises, vanquished,  \\
    && suicidésquad, nuptials &&execution \\
    \cmidrule(lr){1-1} 
    \cmidrule(lr){2-3} \cmidrule(lr){4-5}
    \textit{Movement}  & 
    35\% &  invades, hike, pilgrimage, riding, hiked &
    17\% & re, fled, remove, charge, charges \\
    \cmidrule(lr){1-1} 
    \cmidrule(lr){2-3} \cmidrule(lr){4-5}
    \multirow{2}{*}{\textit{Conflict}}  & 
    \multirow{2}{*}{9\%} &  \multirow{2}{*}{fightnight, fighton, invades, flew, conquered} &
    \multirow{2}{*}{18\%} & assault, raids, warfare, vanquished,  \\
    &&&& clash \\
    \cmidrule(lr){1-1} 
    \cmidrule(lr){2-3} \cmidrule(lr){4-5}
    \textit{Contact}  & 
    17\% &  reunited, graduated, love, stopped, appointments &
    12\% & hacked, signed, join, hearing, talks \\
    \cmidrule(lr){1-1} 
    \cmidrule(lr){2-3} \cmidrule(lr){4-5}
    \multirow{2}{*}{\textit{Personnel}}  & 
    \multirow{2}{*}{7\%} &  electionday2016, graduation, headed, graduated,  &
    \multirow{2}{*}{12\%} & \multirow{2}{*}{resign, deal, quits, signing, resignation} \\
    & & landed \\
    \bottomrule
    \end{tabular}
    \caption{Compared to \texttt{Random} tweets, \texttt{Check-In} tweets contain more special and positive events such as weddings or graduations, detected using \citet{ma2021eventplus}. For each event category, we show the proportion of the event category detected out of all detected events in the \texttt{Check-In} and \texttt{Random} datasets, respectively, as well as the top 5 event triggers ranked by odds ratio~\cite{sun2021men}. We only show event categories that appear more than 10\% of the time combined. }
    \label{tab:events}
\end{table*}

\begin{figure}
    \centering
    \includegraphics[width=\linewidth]{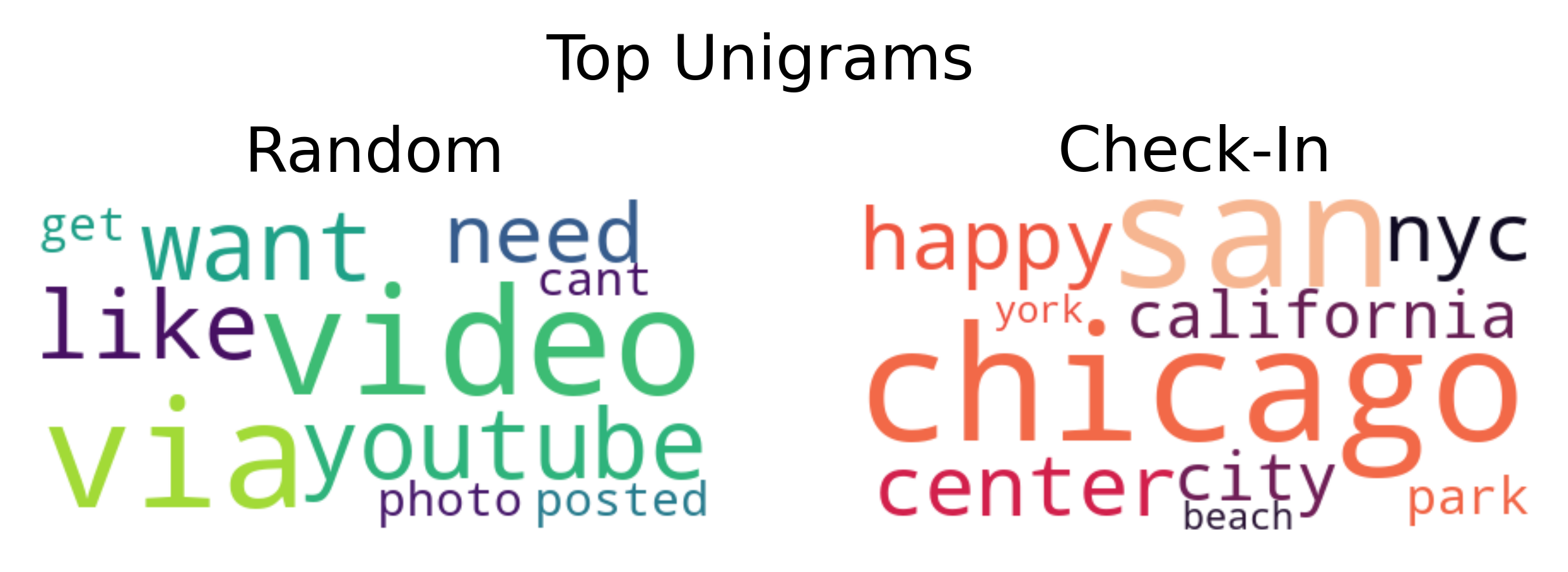}
    \caption{The top 10 unigrams most commonly associated with \texttt{Random} and \texttt{Check-In} tweets using weighted log odds~\cite{monroe2008fightin}. \texttt{Check-In} tweets unigrams reference common tourist locations and tokens that express positivity (``happy''). }
    \label{fig:unigram_wc}
\end{figure}
\begin{figure}
    \centering
    \includegraphics[width=\linewidth]{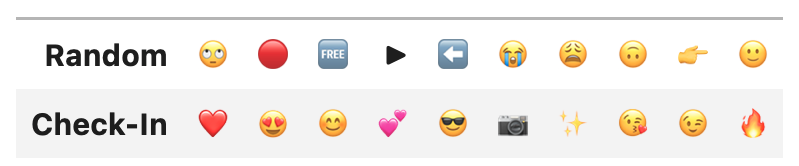}
     \caption{The top 10 emojis most commonly associated with \texttt{Random} and \texttt{Check-In} tweets using weighted log odds~\cite{monroe2008fightin}. \texttt{Check-In} tweets emojis are noticeably more positive. }
    \label{fig:emojis}
\end{figure}
\subsection{RQ2: Differences in Content}

\subsubsection{Check-In Posts Reference Touristy Spots.} The top 10 unigrams most associated with \texttt{Random} and \texttt{Check-In} tweets are shown in Figure \ref{fig:unigram_wc}. These unigrams are selected based on the weighted log odds~\cite{monroe2008fightin}. The unigrams most commonly associated with \texttt{Check-In} contain more references to popular tourist spots such as NYC, Chicago, or the beach. One top unigram of the  \texttt{Check-In} dataset is ``happy'', further indicating the prevalence of positive emotions among \texttt{Check-In} tweets.

\subsubsection{Check-In Posts Contain Positive Emojis.} From Figure \ref{fig:emojis}, we see that the top emojis used in \texttt{Check-In} tweets, as measured by the weighted log odds \cite{monroe2008fightin} are noticeably more positive than those of \texttt{Random} tweets; the former contains happy faces and hearts, while the latter consist of sad, contempt, or neutral emojis.

\begin{figure}
    \centering
    \includegraphics[width=\linewidth]{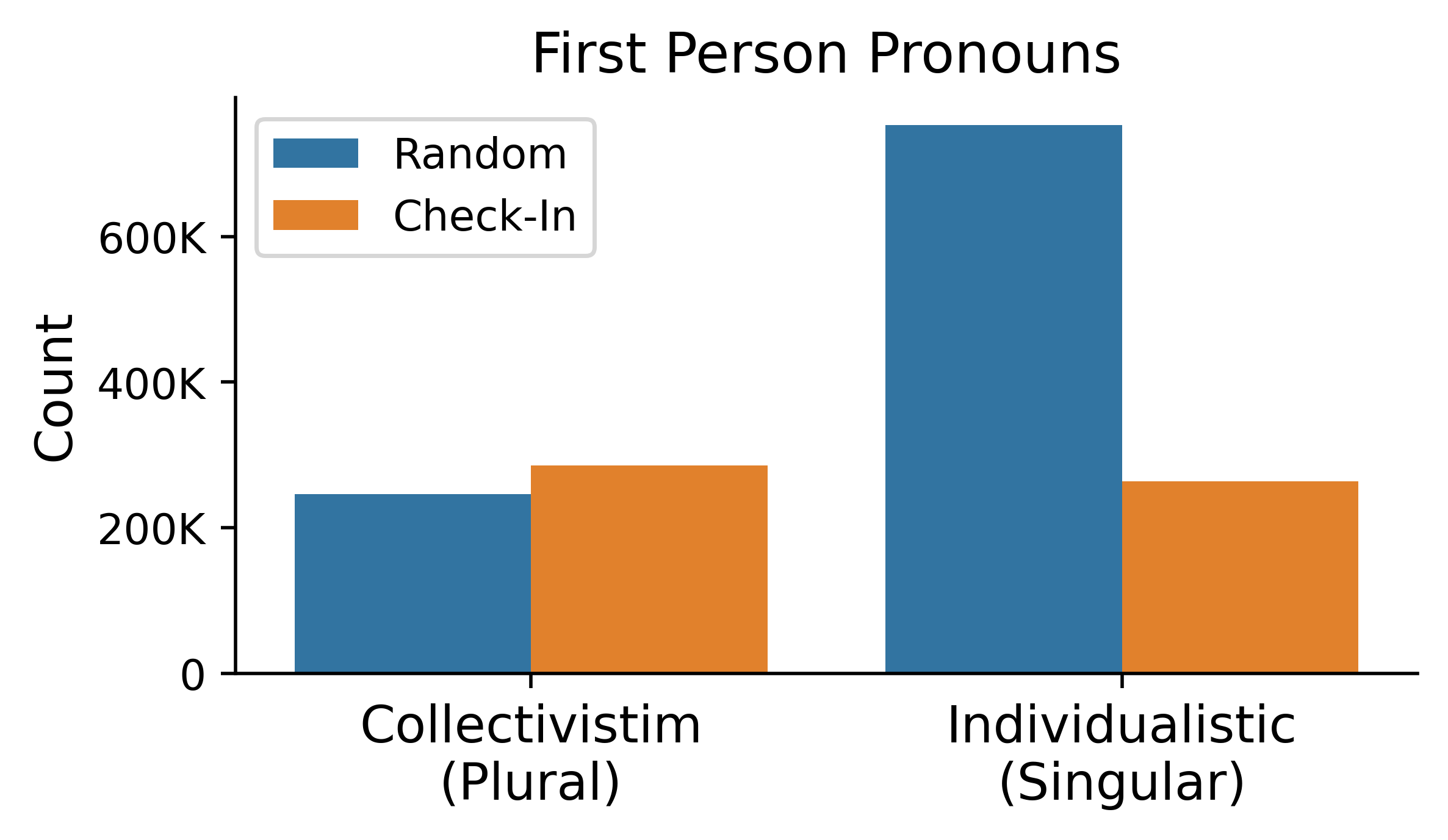}
    \caption{\texttt{Random} tweets employ substantially more first-person pronouns, especially significantly more individualistic pronouns ($z$-test, $p<0.001$).}
    \label{fig:pronouns}
\end{figure}

\subsubsection{Check-In Posts Use More First-Person Plural Pronouns.} \texttt{Check-In} tweets contain half as many first-person singular pronouns. The counts of first-person plural pronouns are comparable, but \texttt{Check-In} tweets still have significantly more first-person plural pronouns ($z$-test $p<.001$). According to \citet{twenge2013changes}, Ttis suggests that \texttt{Check-In} tweets convey a greater sense of collectivism while \texttt{Random} tweets convey a greater sense of individualism. We note that this result contradicts previous work~\cite{karami2021analysis}, which found that check-in posts contain more first-person singular pronouns. One explanation for the lack of first-person singular pronouns is that \texttt{Check-In} users possibly engage in fewer solo activities.

\subsubsection{Check-In Posts Contain Special and Positive Events.}
EventPlus~\cite{ma2021eventplus} extracted events from 7\% and 3\% of the tweets from \texttt{Random} and \texttt{Check-In}, respectively. The most popular event categories are \textit{Life}, \textit{Movement}, and \textit{Conflict}. Table \ref{tab:events} displays the top event triggers for the more frequent event categories. \texttt{Check-In} consists of more positive and also rare experiences such as weddings, adventures, election day, and graduation. There are also keywords such as ``flew'', ``headed'', and ``landed'' that suggest traveling. Even the more hostile-sounding trigger of ``invade" was used in a lighthearted and exuberant manner, such as in the case of ``\textit{2 DAYS until steveaoki invades @OMNIASanDiego}``. Triggers in \texttt{Random} stand in stark contrast with triggers in \texttt{Check-In}, with the former consist of negatively connotated triggers referring to warfare, executions, resignations, and hacking.

\subsubsection{Check-In Images Contain More Objects.}

\begin{figure}
    \centering
    \includegraphics[width=\linewidth]{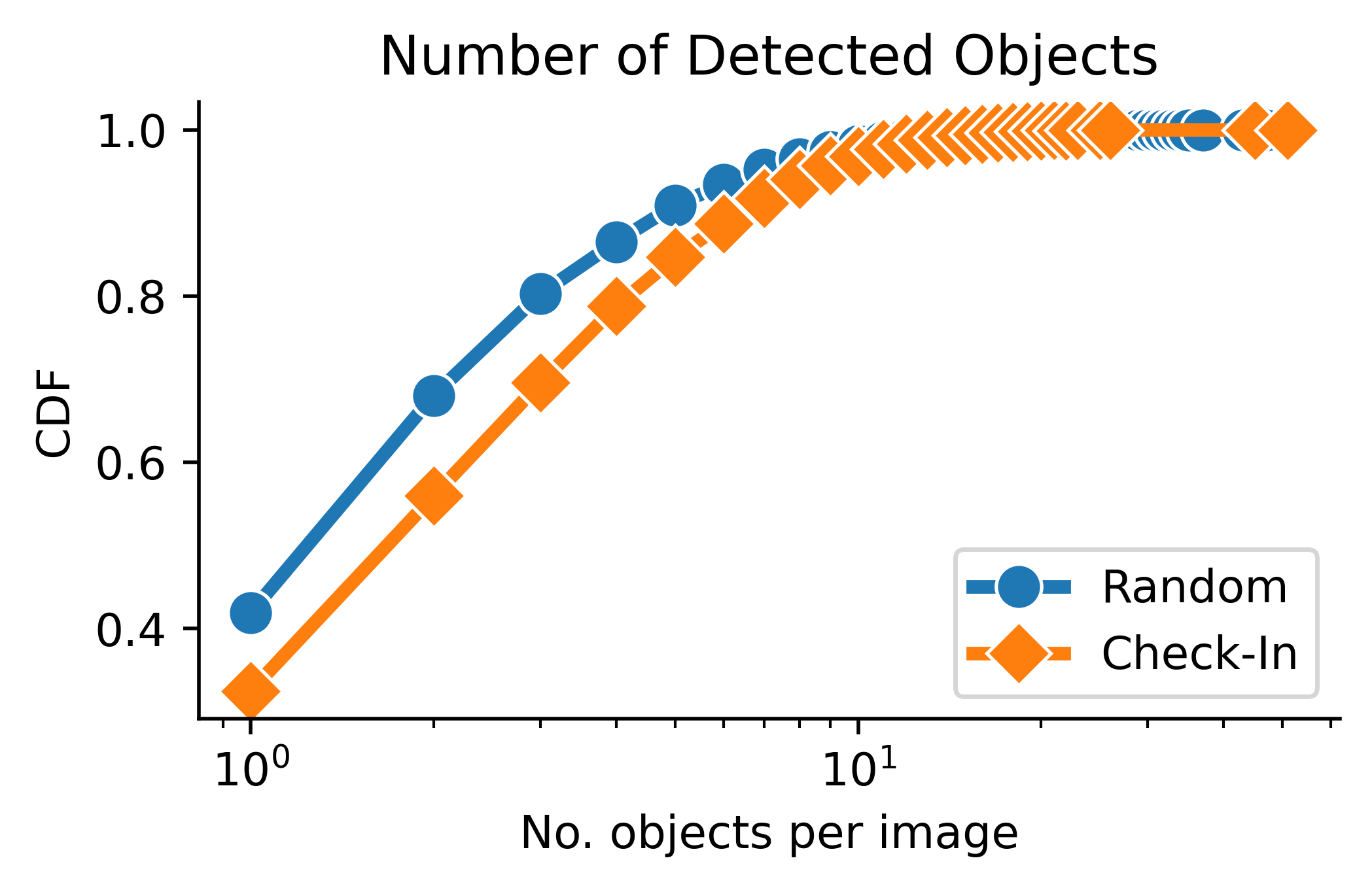}
    \caption{The CDF of the number of detected objects per image using Faster R-CNN~\cite{ren2015faster}. There are significantly more detected objects (Mann-Whitney U-test, $p<0.001$) in \texttt{Check-In} images.}
    \label{fig:n_obj}
\end{figure}

Figure \ref{fig:n_obj} shows the CDF of the number of detected objects per image using Faster R-CNN~\cite{ren2015faster}. We find that \texttt{Check-In} images (mean $=4.0$, median $=3$) have significantly more detected objects than \texttt{Random} images (mean $=3.2$, median $=2$) using a Mann-Whitney U test ($p<0.001$). However, there is no significant difference in the proportion of the types of objects detected ($p>0.05$, Chi-square test). \textit{Persons} are the most frequently detected objects in both datasets, consisting of more than 70\% of all objects detected (see Appendix). These results suggest while \texttt{Check-In} images have more visually distinct components than \texttt{Random} images, the nature of the said components is not characteristically different. 
\begin{figure}
    \centering
    \includegraphics[width=0.75\linewidth]{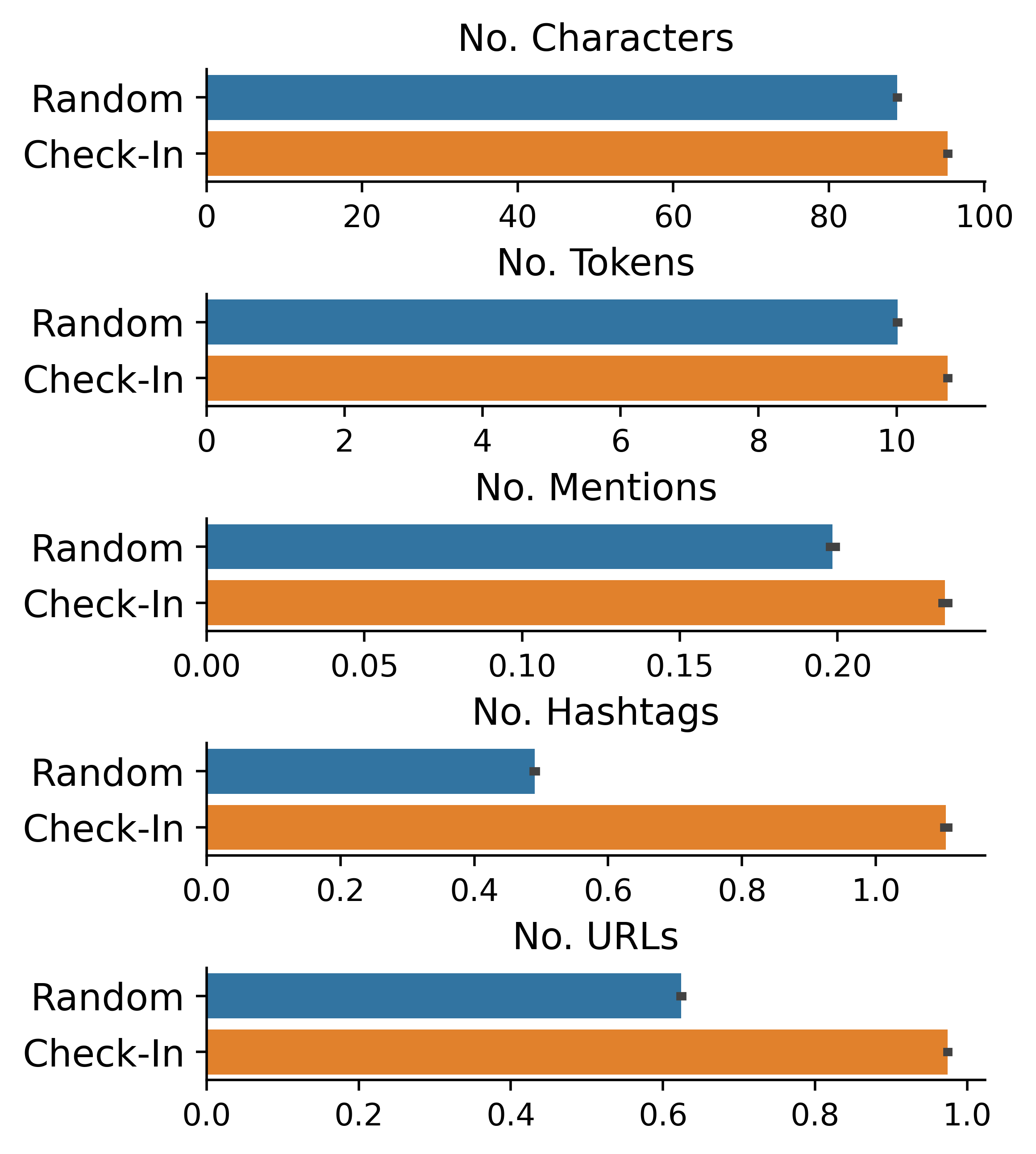}
    \caption{\texttt{Random} and \texttt{Check-In} tweets are differentiated by several auxiliary characteristics. \texttt{Check-In} tweets are longer in length and contain more mentions, hashtags, and URLs (Mann-Whitney U-tests, $p<.001$).}
    \label{fig:meta_characteristics}
\end{figure}

\subsubsection{Check-In Posts Contain More Auxiliary Features.} Figure \ref{fig:meta_characteristics} compares several other auxiliary features between \texttt{Random} and \texttt{Check-In} tweets. \texttt{Check-In} tweets are consistently longer and contain more mentions of other users, hashtags, and URLs. All differences are significant using Mann-Whitney U-tests ($p<0.001$).

\subsection{RQ3: Engagement}
\begin{figure}
    \centering
    \includegraphics[width=\linewidth]{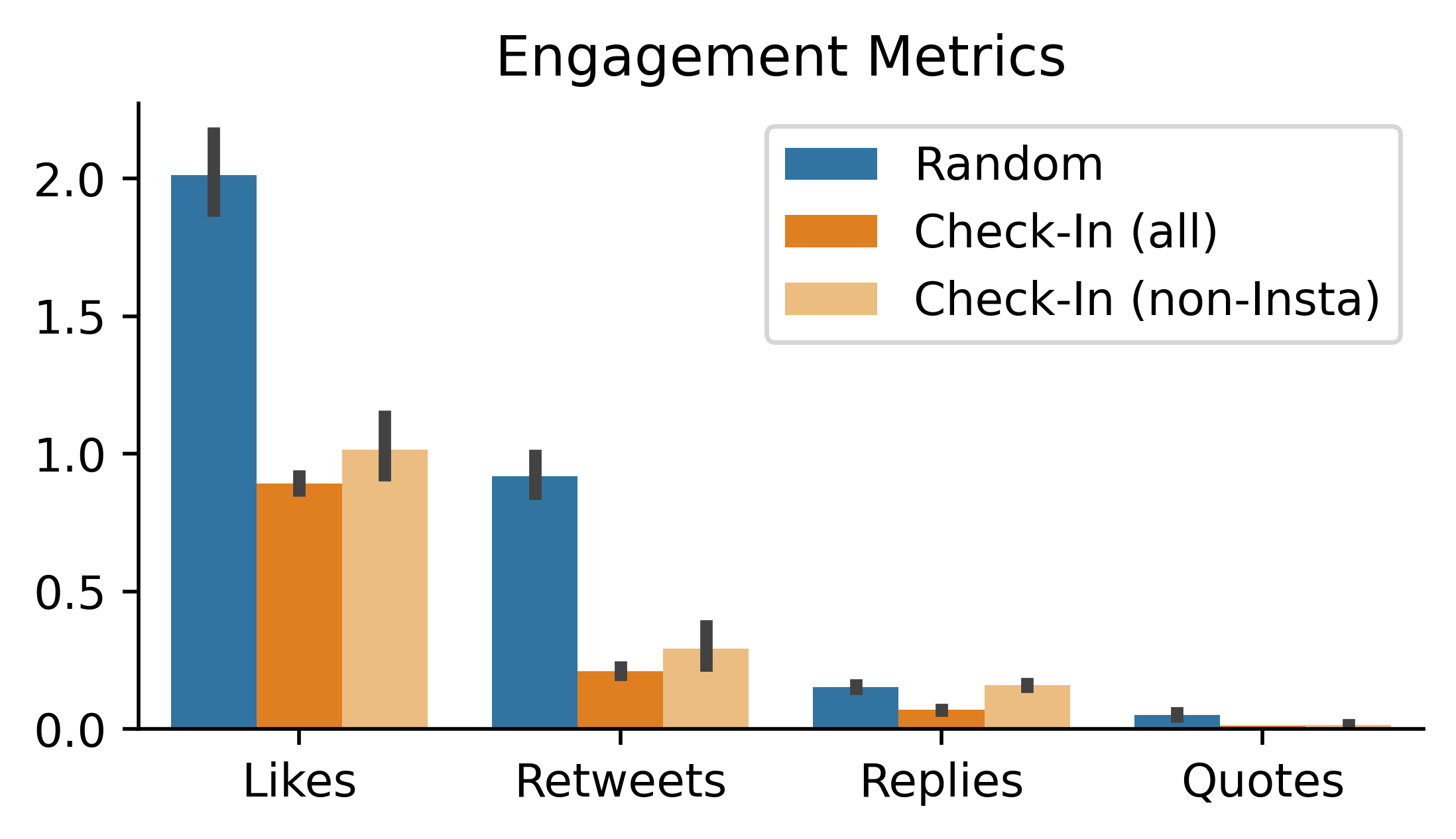}
    \caption{\texttt{Random} tweets rank significantly higher than \texttt{Check-In} tweets across all measures of engagement (Mann-Whitney U-tests, $p<.001$).}
    \label{fig:engagement}
\end{figure}

Figure \ref{fig:engagement} shows the average engagement metrics yielded by \texttt{Random} and \texttt{Check-In} tweets in terms of the number of likes, retweets, replies, and quotes. \texttt{Random} tweets rank substantially higher in engagement, receiving twice as many likes and thrice as many retweets. There is little difference between the engagement of non-Instagram sourced \texttt{Check-In} tweets and the rest of the \texttt{Check-In} tweets. As such, we argue that  \texttt{Check-In} tweets garner significantly less attention compared to \texttt{Random} tweets.

\begin{figure}
    \centering
    \includegraphics[width=\linewidth]{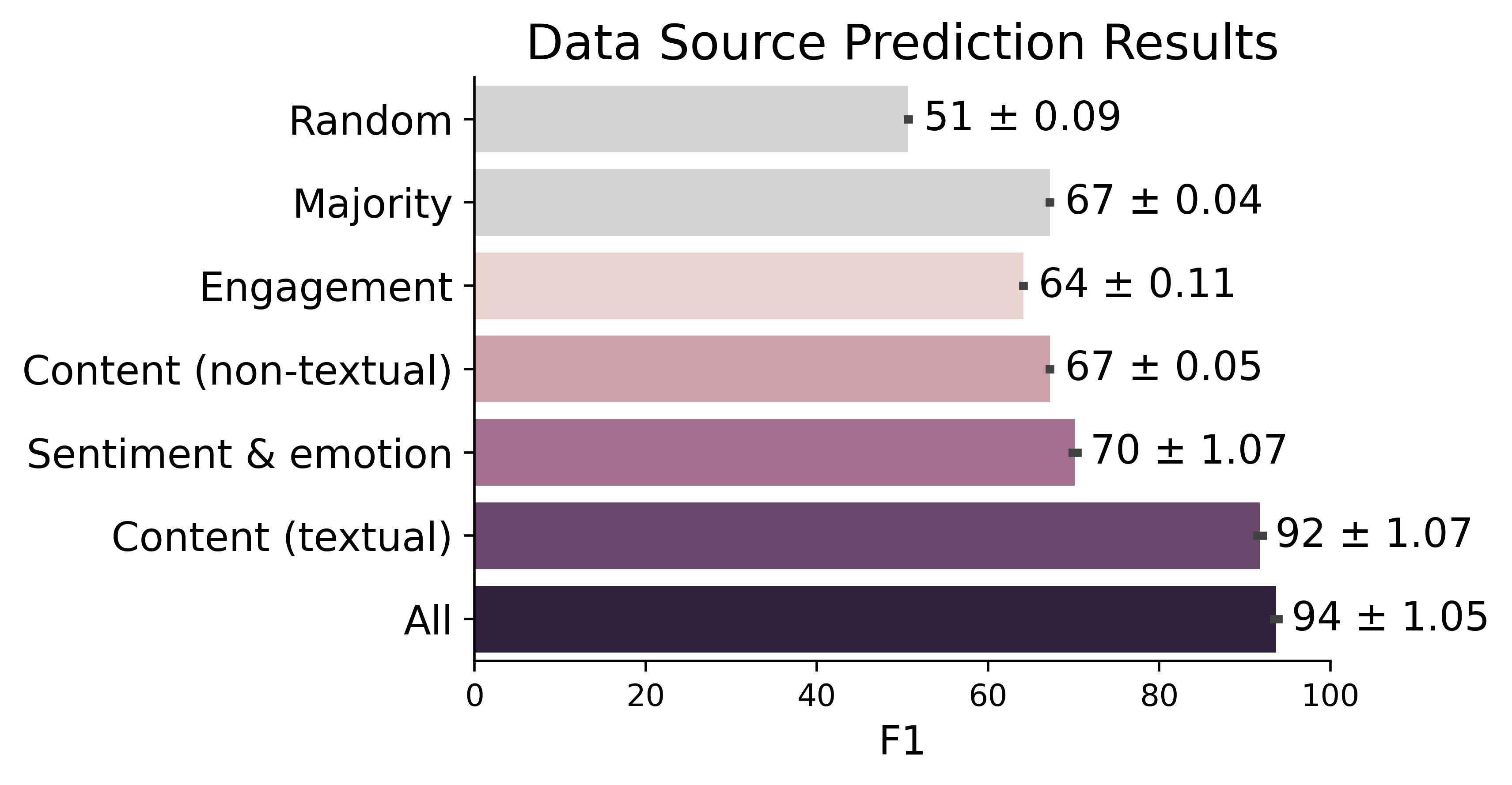}
    \caption{The F1 scores of the binary classification whether a tweet belongs to the \texttt{Check-In} dataset or the \texttt{Random} dataset using various combinations of input features categories. The singularly most informative feature is textual content.}
    \label{fig:prediction}
\end{figure}

\begin{table}
    \centering
    \begin{tabular}{rccc}
        \toprule
         & Neg. & Neu. & Pos.  \\
         \midrule
         All & 91.8 & 95.0 & 94.5\\
         Content (textual) & 88.1 & 92.8 & 91.8\\
         \bottomrule
    \end{tabular}
    \caption{The F1 scores of tweets tagged with negative sentiment is the lowest using the model trained on all features and on textual content features.}
    \label{tab:f1_by_sentiment}
\end{table}
\subsection{Prediction Results}\label{sec:prediction}

We use F1-scores to evaluate the binary classification prediction of  tweet data type, i.e., whether the tweet belongs to the \texttt{Random} class or the \texttt{Check-In}. All scores are determined from the average of 10 repeated runs of five-fold cross-validation. Two additional baselines are used: random prediction based on the distribution of the training labels, and majority prediction based on the majority training label. 

Of all the non-textual feature categories, affect is the most informative predictor of data type. Engagement features are the least predictive, performing slightly worse than the majority baseline. Non-textual media content performs similarly to the majority baseline. Unsurprisingly, using textual features attains a considerably improved F1 score of 92, possibly because most information is encoded in the actual text. Combining textual and non-textual features results in an F1 score of 94, which is a significant improvement ($p<0.001$, $t$-test) over only textual features.

Upon further examination of the predictive performance, we find predictive modeling had the most difficulty distinguishing tweets tagged with negative sentiment (Table \ref{tab:f1_by_sentiment}). While the F1 scores for the tweets with neutral or positive negative were both around 95, the F1 scores of the tweets with negative sentiment is only 91.8. This same pattern is repeated with the model trained only on textual content features, the singularly most predictive feature. These differences suggest that the geolocated negative tweets are more similar to non-geolocated negative tweets, whereas geolocated neutral and positive tweets are more distinct from non-geolocated ones.

We note that we deliberately ignored the source of the tweet in the prediction pipeline since \texttt{Check-In} tweets are overwhelmingly sourced from Instagram. The fact that we can predict with high reliability the data type from other features underscores the substantial differences between tweets that are geotagged and tweets that are not. 

\section{Discussion}

This work aims to provide a deeper understanding of precisely how geotagged social media posts differ from non-geotagged social media posts. Comparing two large samples of check-in and random tweets, We find that check-in posts exhibit significant positivity bias, both in terms of sentiment and emotions. This finding is corroborated by the outsized presence of keywords such as ``happy'', positive emojis, and stereotypically positive events such as weddings or graduations. We also find explicit references to popular tourist locations such as NYC. Therefore, we hypothesize that users most commonly geotag when they are on vacation, participating in exciting events, or generally out-and-about. Such behavior could be attributed to social motivations for using geolocation~\cite{lindqvist2011im,tasse2017state,patil2012check,patil2012reasons,guha2013can}. 

Several other characteristics further distinguish check-in from non-check-in tweets. We find that check-in tweets contain significantly more textual and non-textual features. They are also packed with more auxiliary features such as hashtags, mentions, or URLs. In addition, the attached images of check-in tweets contain significantly more objects. In terms of pronouns, check-in tweets boast less individualistic pronouns (`I' pronouns) than collectivistic pronouns (`We' pronouns) \cite{twenge2013changes}. 

Check-in tweets also have lower engagement, evidenced by lower numbers of retweets, quotes, replies, and likes. One explanation is that check-in users are less motivated by perceived influence or popularity on Twitter as measured strictly by the engagement metrics we consider. We note that this does not necessarily contradict the hypothesis of socially motivated use of geotags that we discussed earlier, since the majority of check-in posts are posted to Twitter via Instagram, meaning the primary intended medium of their post is Instagram and not Twitter. Moreover, engagement metrics were the worst predictors in distinguishing \texttt{Check-In} tweets from \texttt{Random} tweets, in fact performing slightly worse than the majority baseline, which suggests that the engagement levels are not reliable indicators of whether a post is geotagged. This, however, is still an indication that users who geotag on Twitter are motivated by distinct reasons. Since data from Instagram is challenging to obtain, the exploration of geotagging and engagement on Twitter versus Instagram is out of the scope of the work.

\subsection{Recommendations}

Though only an estimated less than 1\% of tweets are geotagged on Twitter~\cite{ajao2015survey}, geolocated tweets remain an invaluable resource to study a myriad of real-world phenomena on Twitter~\cite{de2009omg,crooks2013earthquake,gupta2013faking,padmanabhan2014flumapper,botta2015quantifying,jurdak2015understanding,hasan2016understanding,yaqub2018analysis,dahal2019topic,hobbs2019effects,phillips2021social,wang2022global}, necessitating the need to understand how check-in posts differ from regular posts. Our findings imply that geotagged data are fundamentally different from other, non-geolocated data. We now discuss several concrete recommendations.

For human mobility research, it is important to note that geotagged social media posts are optional  disclosures that are typically about situated, special, personal, and overwhelmingly positive events. Popular examples of geotagged tweets include occasions such as weddings, graduations, or vacation check-ins. The more quotidian examples of commutes and routine travels will likely not be geotagged. As a result, studies of mobility based on geolocation data only capture the share of people who travel for vacations or special occasions. Moreover, geotagged posts are distinct from other movement data such as phone service location in that they are typically opt-in, confounding research on general human mobility using geolocated social media data. Improvement in this area would require alternative tools to infer location or mobility based on other modalities such as texts or photos on social media or wifi and cell services. 

The results of our work could strengthen research on measuring public attitudes and well-being via geolocation by understanding what makes them unique, offering ways to generalize their research to the wider public. First, users who geotag are motivated to use social media for reasons different from the usual users \cite{lindqvist2011im,patil2012check,patil2012reasons,guha2013can,tasse2017state}. Second, since geotagged posts are overwhelmingly positive, evaluating only geotagged posts would perhaps overestimate the true positivity of public opinions. While some may argue that although geolocated data is not representative of all social media data, they are comparable within themselves. For example, we can compare the state-wide public responses to a policy with geotagged tweets of Texas and New York. However, our work also implies that users geotag more frequently at the location they travel to for pleasure, potentially over-inflating the positivity of popular tourist destinations. As a result, we advise researchers to devise methods to filter out geolocated posts by users who are not at their home location. By recognizing the differences and limitations of geolocated datasets, researchers can better avoid the pitfalls of data bias.

That being said, though our work shows most geotagged posts are positive, we believe there is still tremendous value and potential in analyzing geotagged posts to study typically negative events such as natural disasters \cite{crooks2013earthquake,de2009omg,gupta2013faking}, disease outbreaks \cite{padmanabhan2014flumapper}, or civil unrest \cite{larson2019social}, provided they include additional keyword filters on top of geotagging to remove the usual geotagged posts.

For advertisers, our work implies that users who geotag are likely in a positive mood vacationing or attending rare and special events. They could therefore adjust their promotion strategies to better cater to people who are on the go or otherwise not taking part in their usual routine.

\subsection{Broader Implications}
Though the geotagging feature is no longer a prominent feature of Twitter \cite{niemanlab}, location sharing overall is on the rise on other platforms such as Instagram, Snapchat, and TikTok, becoming a ubiquitous feature in online social networks. By filling the void of research in understanding the nature of geolocated posts, our work carries broader implications for social media research overall.

\subsection{Limitations} \label{sec:limitations}
There are several data bias limitations to consider. First, since our \texttt{Check-In} mostly originated from Instagram, one concern is that the conclusions we draw are as much about differences between Instagram posts and Twitter posts as they are about check-in posts and non-check-in posts. However, we believe this does not change the fact that users who use geolocation on Twitter via Instagram are \textit{also} plainly users who use geolocation on Twitter. Additionally, our collection method of a random sample of tweets rests on the assumption that tweets are relatively temporally randomly distributed. Our \texttt{Check-In} dataset is also limited to 30 major US and Canadian cities, meaning the dataset could be geographically biased. Some posts could be spam, though we believe the large sample size of our datasets offsets the confounds of spam posts.  Another data bias is that only half of the \texttt{Random} data comes from mobile sources, while \texttt{Check-In} data is exclusively mobile. That said, a previous study found that mobile tweets are in fact 25\% more negative than web-based tweets \cite{murthy2015we}, which further underscores the positivity of mobile-only \texttt{Check-in} tweets. Lastly, our analysis of attached media is limited in its capacity to explain media differences between check-in and non-check-in posts because we were unable to obtain images posted to Instagram, which made up the bulk of \texttt{Check-In} attached media.

\section{Conclusion}
Despite the prolific use of geolocated data to solve social science problems such as human mobility, it is known that users who use geolocation are not representative of the population. Less is known about \textit{how} geolocated data are different from non-geolocated data. To bridge this gap in research, we examine the characteristics that differentiate check-in (geotagged) tweets on Twitter/Instagram from a random dataset of tweets. We find that check-in posts demonstrate considerably more positive sentiment and emotions, include more positive emojis, and contain more explicit references to rare and positive occasions (e.g., weddings) as well as touristy spots (e.g., NYC). They also use more first-person plural pronouns, conveying a higher sense of collectivism. In terms of non-textual content, check-in posts consist of more features such as hashtags, mentions, or more objects in their attached images. Another differentiating aspect is that check-in posts garner substantially less engagement in terms of retweets, likes, etc. Our findings have bear important  recommendations for future studies utilizing geolocation data. 
\section*{Ethical Statement}
Of paramount importance in any research concerning user data is to protect user privacy, user identity, and malicious use of the data. This is especially crucial concerning data with sensitive location information such as ours. Throughout this study, we do not attempt to surmise the mobility of geolocated users nor utilize the provided location of check-in tweets beyond the fact that they \textit{are} check-in tweets. Additionally, to protect user privacy as well as to adhere to Twitter's data sharing policies, we will only share the data to the extent of the tweet IDs. Any reproduction efforts have to be done by manually hydrating the tweets using one's own approved Twitter API credentials. 

\smallskip\noindent Note: this project was approved by our institution's IRB.

\begin{acks}
The authors thank Leonardo Neves, Maarten W. Bos, Neil Shah, and Yozen Liu for their helpful inputs and insights in the initial stages of this research. The research is partially supported by DARPA (award number HR001121C0169).
\end{acks}
\bibliographystyle{ACM-Reference-Format}
\bibliography{main}

\appendix
\section{Appendix}
\renewcommand{\thefigure}{A\arabic{figure}}
\setcounter{figure}{0}

\subsection{Data Source}
\subsubsection{Mobile Sources.} We operationalize mobile sources as sources that must be used on a phone or tablet. Out of the 14,755 unique sources we designate Instagram, Foursquare, and Untapped as mobile sources. We also designate any sources that contain the strings `phone', `android', `ipad', `blackberry', `tablet', or `ios' as mobile sources. However, this list is not exhaustive and there may be other sources that also represent mobile devices. Thus our estimation of mobile sources is a lower bound. 

\subsubsection{Top Data Sources.}
\begin{figure}[h]
    \centering
    \includegraphics[width=\linewidth]{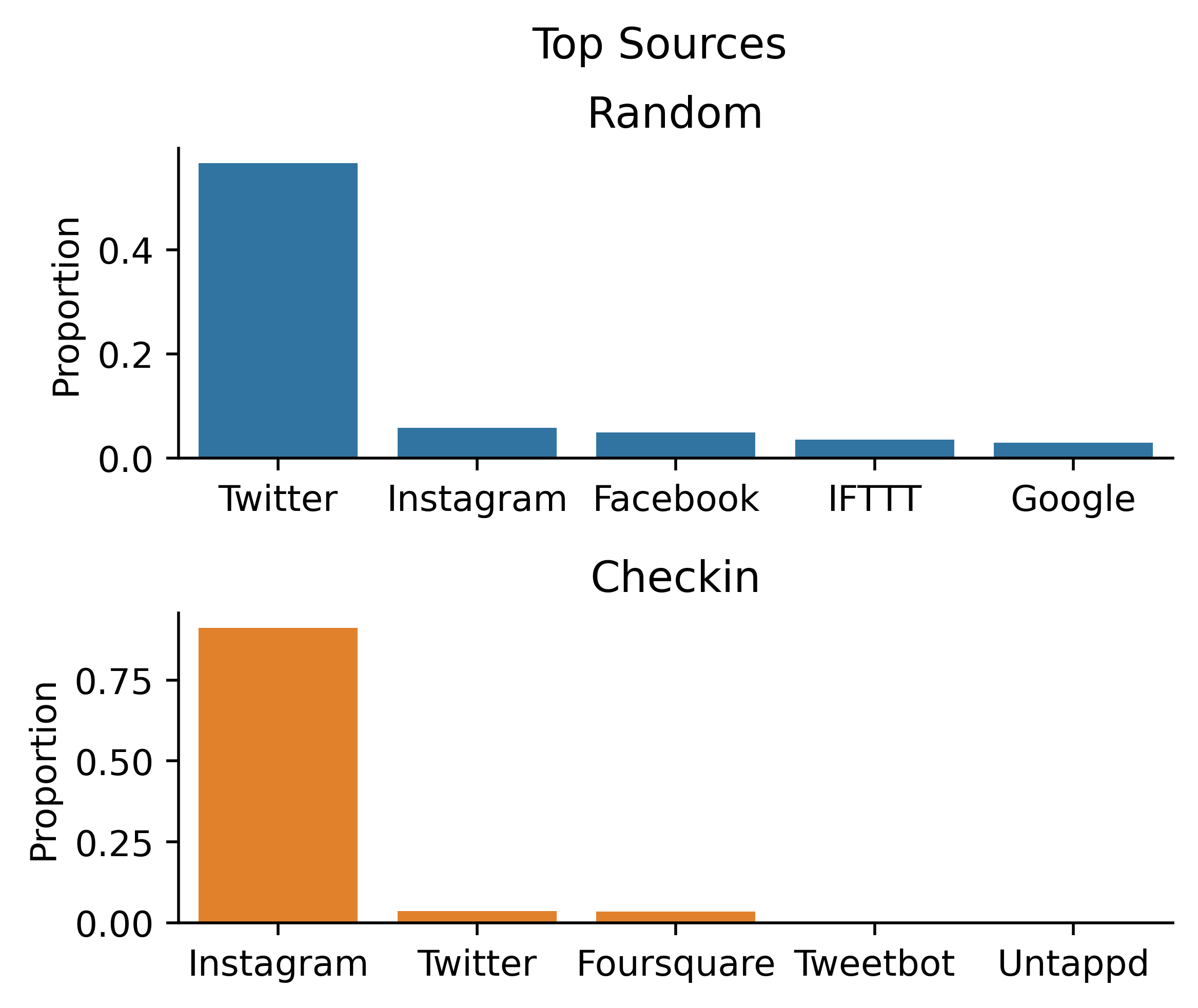}
    \caption{The top five data sources of \texttt{Random} and \texttt{Check-In} tweets.}
    \label{fig:source}
\end{figure}

Figure \ref{fig:source} displays the top five data sources of \texttt{Random} and \texttt{Check-In} tweets, respectively. 

\subsection{Model Parameters \& Cutoff Tuning}
\subsubsection{Hyperparameter Tuning.} The hyperparameter grid we search over are as follows: \texttt{num\_layers}: \{2, 4, 16\}, \texttt{hidden\_dim}: \{32, 168, 768\}, \texttt{learning\_rate}: \{$1\times 10^{-3}, 1\times 10^{-5}, 1\times 10^{-7}\}$, \texttt{l2\_regularization}:  \{$1\times 10^{-2}, 1\times 10^{-3}, 1\times 10^{-5}\}$, \texttt{dropout}: \{0.1, 0.2, 0.5\}. We fix the hidden activation to be ReLU, output activation to be sigmoid, and the optimizer to be Adam. Layer normalization is applied after the input layer and dropout layers are applied after every hidden layer. The batch size is 4,096.

\subsubsection{Faster R-CNN Objects.} 
\begin{figure}[h]
    \centering
    \includegraphics[width=\linewidth]{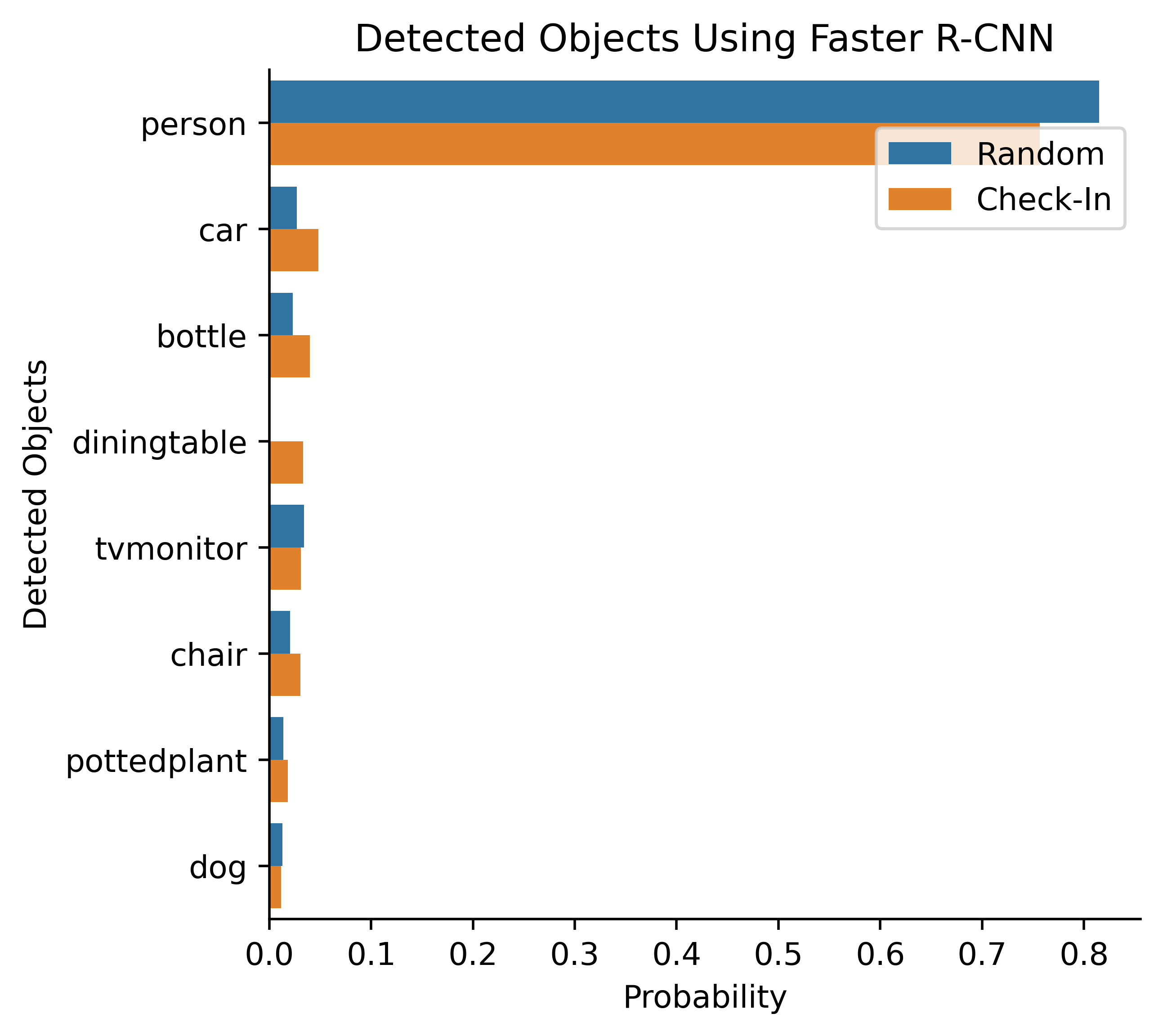}
    \caption{The distribution of objects detected using Faster R-CNN~\cite{ren2015faster} using a 0.95 score cutoff for images in the \texttt{Check-In} and \texttt{Random} dataset. Only objects with more than 1\% total detected occurrences are shown.}
    \label{fig:faster_rcnn_objects}
\end{figure}

We display the distribution of detected objects in Figure \ref{fig:faster_rcnn_objects}, applying the 0.95 cutoff (i.e., only objects with scores greater than 0.95 are included). This is due to two reasons. First, 0.95 is the 90th percentile of all scores. Second, upon manual examination of 20 random images each from \texttt{Check-In} and \texttt{Random}, we find that using 0.95 as the cutoff yields the best detection results. In both datasets, the predominantly detected object is \textit{persons}, followed by \textit{cars} and \textit{bottles}. 65\% of \texttt{Check-In} images and 57\% of \texttt{Random} images have at least one detected objects.




\end{document}